\documentclass[journal,twoside,web]{ieeecolor}
\usepackage{generic}
\usepackage{cite}
\usepackage{amsmath,amssymb,amsfonts}
\usepackage{algorithmic}
\usepackage{graphicx}
\usepackage{textcomp}

\usepackage{algorithm}
\usepackage{booktabs}
\usepackage{comment}

\usepackage[colorlinks,urlcolor=black,linkcolor=black]{hyperref}

\begin{document}

\title{Exploiting Hierarchical Interactions for Protein Surface Learning}
\author{Yiqun Lin, Liang Pan, Yi Li, Ziwei Liu, and Xiaomeng Li$^*$
\thanks{Manuscript received xx/xx, 2023.}
\thanks{$^*$: Xiaomeng Li is the corresponding author.}
\thanks{Yiqun Lin, Yi Li, and Xiaomeng Li are with the Department of Electronic and Computer Engineering, Hong Kong University of Science and Technology, Hong Kong SAR (e-mail: \{ylindw, ylini, eexmli\}@ust.hk).}
\thanks{Liang Pan and Ziwei Liu are with the School of Computer Science and Engineering, Nanyang Technological University, Singapore (e-mail: \{liang.pan, ziwei.liu\}@ntu.edu.sg).}
}

\maketitle

\newcommand{\nickname}{HCGNet}
\newcommand{\etal}{\textit{et al.}}

\newcommand{\ud}[1]{{#1}}

\begin{abstract}

Predicting interactions between proteins is one of the most important yet challenging problems in structural bioinformatics.
Intrinsically, potential function sites in protein surfaces are determined by both geometric and chemical features.
However, existing works only consider handcrafted or individually learned chemical features from the atom type and extract geometric features independently.
Here, we identify two key properties of effective protein surface learning:
1) relationship among atoms: atoms are linked with each other by covalent bonds to form biomolecules instead of appearing alone, leading to the significance of modeling the relationship among atoms in chemical feature learning.
2) hierarchical feature interaction: the neighboring residue effect validates the significance of hierarchical feature interaction among atoms and between surface points and atoms (or residues).
In this paper, we present a principled framework \ud{based on deep learning techniques}, 
namely Hierarchical Chemical and Geometric Feature Interaction Network (\nickname{}), for protein surface analysis by bridging chemical and geometric features with hierarchical interactions.
Extensive experiments demonstrate that our method outperforms the prior state-of-the-art method by 2.3\% in site prediction task and 3.2\% in interaction matching task, respectively.
\ud{Our code is available at} \href{https://github.com/xmed-lab/HCGNet}{https://github.com/xmed-lab/HCGNet}.

\end{abstract}


\section{Introduction} \label{sec:introduction}
\IEEEPARstart{P}{roteins} composed of amino acids are large, complex molecules that play critical roles in all living organisms. Proteins can provide various functions with different structures for organisms, including causing biomedical reactions, acting as messengers, and balancing fluid. From a biological perspective, proteins can be described in four structural levels, including primary, secondary, tertiary, and quaternary structures. The amino acid sequence (primary) determines the three-dimensional (3D) structure (tertiary) of the protein, and its functions mainly depend on its 3D structure. Therefore, exploring the functions of different proteins' 3D structures is crucial for understanding their working mechanism, which benefits many applications, such as new drug development~\cite{shen2020machine}.

\begin{figure}[t]
\centering
\includegraphics[width=0.95\linewidth]{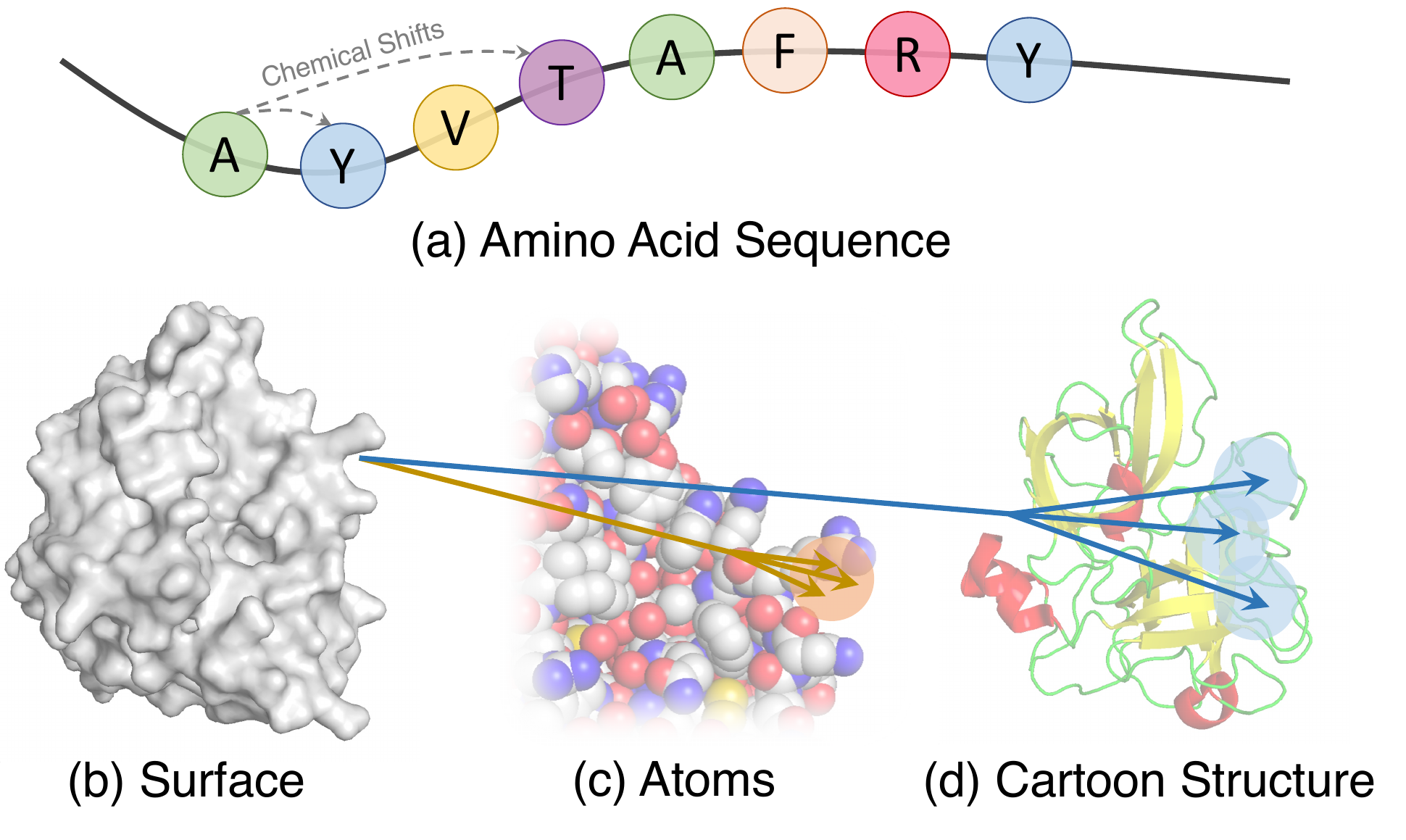}
\vspace{-3mm}
\caption{(a-d) show a protein's amino acid sequence, surface, atoms, and cartoon structure (a simplified representation based on the secondary structure), respectively. (a) Due to the neighboring residue effect~\cite{wang2002investigation}, multiscale relationships among atoms and between surface points and atoms should be considered in protein function analysis. (b-d) Our key idea is to model the hierarchical feature interactions between chemical (atom/residue) and geometric (surface) features for efficient protein surface learning.}
\label{fig:teaser}
\end{figure}

Proteins carry out their functions by interacting with other proteins or molecules.
Therefore, predicting protein-protein or protein-biomolecule \ud{interaction (e.g., interaction sites and interaction matching)} becomes one of the most important and challenging problems in structural bioinformatics. 
Proteins often associate through hydrophobic patches on their surfaces. Amino acid residues in the interaction interfaces can give a general indication of hydrophobicity~\cite{jones1996principles}. Various studies about the electrostatic nature of the protein-protein interface demonstrate that the associating surfaces interacting with one another have charge complementarity~\cite{novotny1992electrostatic,roberts1991electrostatic} or electrostatic complementarity~\cite{braden1995structural,demchuk1994receptor,vakser1994hydrophobic}. In addition, the size and shape of protein interfaces, measured with solvent accessible surface area ($\Delta \text{ASA}$)~\cite{lee1971interpretation}, varies in different types of protein complexes. Earlier work~\cite{jones1996principles} shows that the range of $\Delta \text{ASA}$ in the heterocomplexes is smaller than in the homodimers, and the larger molecules usually have larger interfaces. The interaction between proteins also relies on shape complementarity~\cite{lawrence1993shape}, such as convex bulges and concave pockets. Therefore, whether proteins interact with other molecules and where the interaction sites are located depends on both chemical features and 3D geometric shapes.

Existing research works on protein interaction prediction~\cite{masif_nature_method,mai2020multiscale,rego2021identifying,vakser1994hydrophobic} utilize handcrafted chemical features \ud{(e.g., hydrophobicity and charges)} and adopt deep neural networks to learn geometric features \ud{(e.g., curvatures and normals)} from 3D structures. 
Recently, dMaSIF~\cite{dmasif_cvpr2021} is proposed to encode the atom type into a one-hot vector followed by multi-layer perceptrons (MLPs) to learn chemical features. Their experiments show that learned chemical features can perform as well as or better than handcrafted ones.

Nonetheless, dMaSIF~\cite{dmasif_cvpr2021} has two main limitations. 
Firstly, it does not consider the relationship among atoms. The chemical feature of each atom is learned individually only from the atom type. However, in a molecule, atoms are linked together by covalent bonds to form functional groups that cause chemical reactions of molecules~\cite{mcnaught1997compendium}. This indicates that the chemical feature is not determined by a single atom, but by a group of atoms linked to each other.
Secondly, it ignores hierarchical feature interactions between surface points and atoms. For each surface point, only the nearest 16 atoms around the surface point are gathered as the chemical feature. However, the neighboring residue effect~\cite{wang2002investigation} points out that amino acid residues can bring chemical shifts to their neighboring residues; see Figure~\ref{fig:teaser}.a. This reveals that the chemical properties of the protein surface are not only determined by atoms in the residues near the surface but also rely on atoms in some residues far away from the surface. The same goes for atoms in chemical feature learning.

To this end, we propose a novel 
\textbf{H}ierarchical \textbf{C}hemical and \textbf{G}eometric Feature {I}nteraction \textbf{Net}work 
(\textbf{\nickname{}}) 
to learn both chemical and geometric features in a hierarchical and interactive way. Our key idea is to capture the multiscale relationship among atoms to learn chemical features and hierarchically model their interactions with geometric features. Specifically, we introduce a dual hierarchical framework to extract chemical features and geometric features, respectively. By designing a chemical feature propagation module, the hierarchical chemical features from atoms (or residues) can be propagated to surface points for protein surface learning; see Figure~\ref{fig:teaser}.b-d. Our experiments show that \nickname{} outperforms prior state-of-the-art (SoTA) methods in protein surface learning. To summarize, the main contributions are as follows:
\begin{itemize}
    \item We highlight the importance of hierarchical relationships among atoms and between chemical and geometric features, which are new insights for protein surface learning and have been overlooked in previous works. 
    \item We propose \nickname{} to model hierarchical interactions in chemical feature learning and between chemical and geometric features for effective protein surface analysis. 
    \item Our method outperforms prior SoTA method by 2.3\% in site prediction task and 3.2\% in interaction matching task. \nickname{} is flexible and has the potential to be used in other protein-biomolecule interaction tasks, such as protein-ligand and protein-DNA/RNA. 
\end{itemize}

\section{Related Work}
\noindent
Discovering the properties and functions of proteins at different structural levels will bring significant benefits to protein design and engineering. By leveraging large public protein databases, biological scientists are able to analyze proteins for a variety of downstream applications, for instance, protein structure prediction~\cite{AlphaFold2021,senior2019protein,senior2020improved,xu2019distance}, protein-protein interaction~\cite{dai2021protein,masif_nature_method,jamasb2021deep,rego2021identifying,dmasif_cvpr2021}, ligand binding affinity~\cite{somnath2021multi,xiong2021featurization}, and protein docking~\cite{masif_nature_method,pierce2011accelerating,renaud2021deeprank,schneidman2005patchdock}.

\vspace{3pt}
\noindent \textbf{Feature learning on protein representation.} 
The amino acid sequence is the simplest way to represent a protein. A straightforward way to process protein sequences is to directly apply techniques in natural language processing, including Word2Vec~\cite{mikolov2013efficient} and Doc2Vec~\cite{le2014distributed}. Recently, UniRep~\cite{alley2019unified} adopted multiplicative LSTMs~\cite{krause2017multiplicative} to learn features for each amino acid residue and averaged all residue features into a representative vector. However, sequence-based methods can not handle the detailed analysis of protein structures due to missing spatial information.

The three-dimensional (3D) structures of proteins are more important than sequences in protein function prediction for rich structural details. Atoms can be treated as points with chemical properties in 3D space. Compared with regular 1D sequences and 2D images, 3D representation learning has difficulties like spatial sparsity and disorder of indexing. Recent works~\cite{simonovsky2020deeplytough,townshend2019endtoend} are proposed to voxelize atom points into regular grids and adopt 3D convolution networks for analysis. However, voxel-based methods still suffer from high computational costs and limited resolution, resulting in insufficient analysis of structural details.
Another effective way is regarding the 3D protein structure as a graph composed of vertices (atoms) and edges (chemical bonds). Fout A., \etal~\cite{fout2017protein} handcraft features — distance and angle for edges, protrusion index, and residue depth for vertices, and utilize graph convolution networks, leading to better performance than SVM-based methods in interaction prediction. HOLOPROT~\cite{somnath2021multi} jointly learns protein representation from both amino acid sequences and atom graphs to capture finer details.

In addition to amino acid sequences and atom graphs, researchers~\cite{masif_nature_method,dmasif_cvpr2021} propose that surface can be a more natural way to analyze protein's functions because surfaces contain both chemical and geometric features, revealing information about protein interactions. Although they introduce chemical features to surface points in a handcrafting or learning way, neither of these works considers the neighboring residue effect and the multiscale relationship among atoms.

\vspace{3pt}
\noindent \textbf{Deep learning on point clouds.} Point clouds have become a more and more popular form of 3D data with the development of sensing devices and diverse learning techniques. To solve the problem of irregularity and disorder, Charles \etal~\cite{qi2016pointnet} propose PointNet to apply shared MLPs on each point and aggregate features with a global max-pooling. PointNet can also be used as a local operator for hierarchical processing~\cite{qi2017pointnetplusplus}, and a lot of works~\cite{li2018pointcnn,thomas2019KPConv,wu2018pointconv} followed, developing various point convolution operators. Moreover, kNN graphs can be constructed from point clouds, and graph convolution networks can be applied for feature learning~\cite{li2019deepgcns,dgcnn}. Recent works expand a local patch of point clouds to an estimated~\cite{tatarchenko2018tangent} or a learned~\cite{lin2020fpconv} 2D plane, and then regular 2D convolutions can be further adopted. 

Although protein atoms can also be represented by a set of points, there are three main differences compared to general point clouds. \textit{a.)}~Points are regarded as homogeneous particles continuously distributed in the real world. Due to the limitation of sensing devices, points are collected discretely, with noise, and sometimes assigned with colors or reflection intensity. Atoms are indeed discretely distributed, free of noise, and have individual chemical properties and radii; \textit{b.)}~Points are captured from object surfaces, and there are almost no points inside, while atoms are distributed not only on protein surfaces but also inside the protein; \textit{c.)}~Geometric features play a decisive role in general point cloud analysis, while chemical features are as important as geometric features in protein structure analysis. Inspired by the above observations, we propose a novel learning architecture to bridge chemical and geometric features with hierarchical interactions, leading to SoTA performance in various protein analysis tasks.

\begin{figure}[t]
\centering
\includegraphics[width=0.95\linewidth]{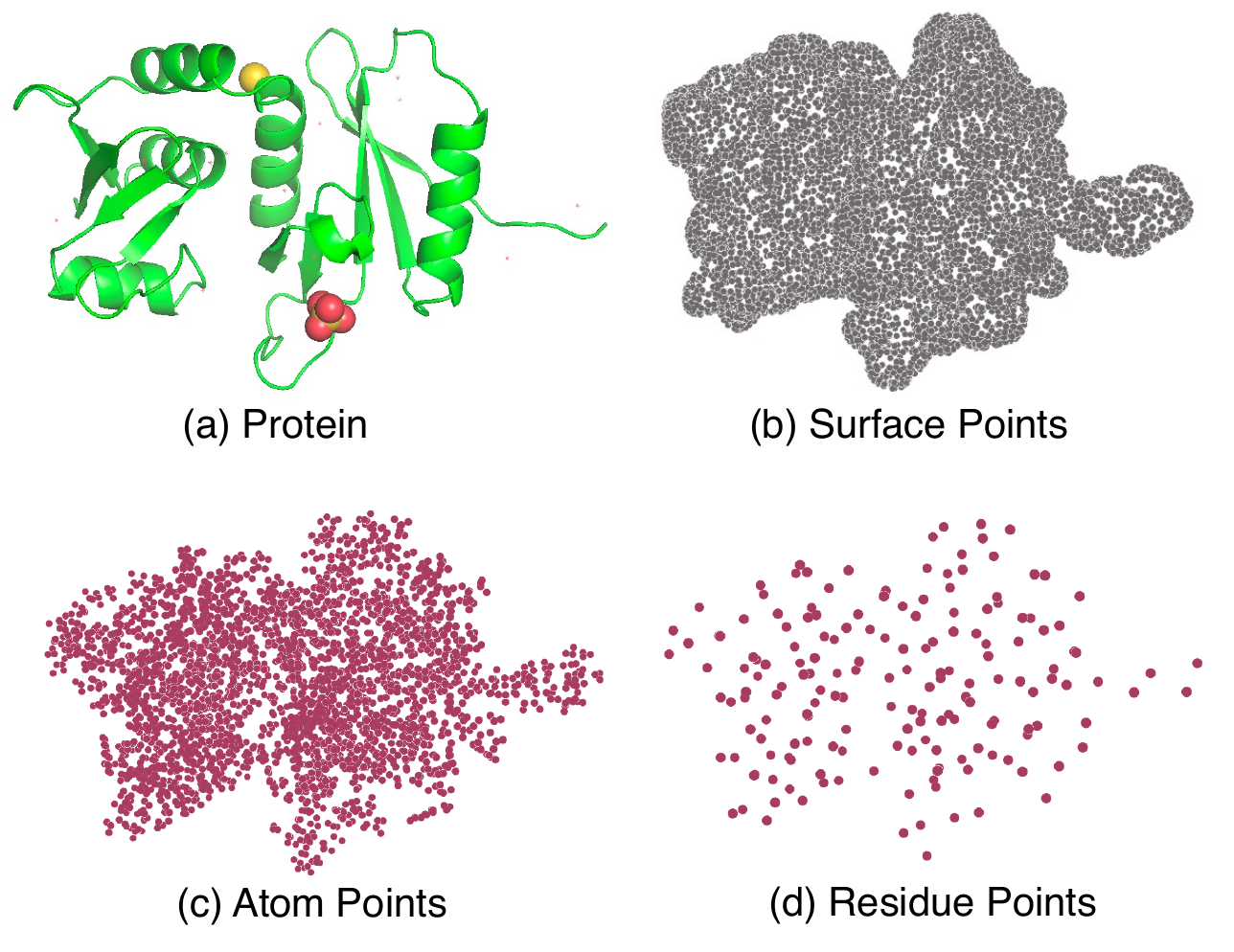}
\vspace{-2.7mm}
\caption{A protein can be represented by three pointsets, including surface points for geometric shape analysis, and atom and residue points for chemical property analysis.}
\label{fig:data_vis}
\end{figure}

\section{Our Approach}
\noindent
In this section, we first introduce protein data, then review some preliminaries about point cloud network design~\cite{qi2017pointnetplusplus}, and finally describe our proposed hierarchical chemical and geometric feature interaction network (\nickname{}).

\subsection{Protein Representation}
\noindent
Proteins are polymers composed of amino acid residues, which are formed by atoms. In the Protein Data Bank (PDB~\cite{berman2003announcing}), each residue is recorded with the amino acid type and a list of atoms, and each atom contains the spatial coordinates and atom type. In the preprocessing, each protein is generated into three pointsets, including atom, residue, and surface pointsets. A visual example is shown in Figure~\ref{fig:data_vis}.

\begin{figure}[t]
\centering
\includegraphics[width=0.9\linewidth]{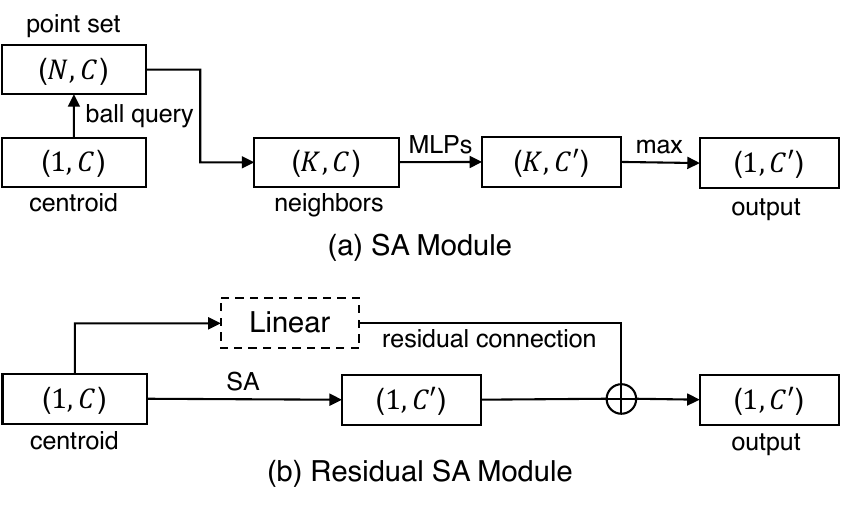}
\vspace{-3mm}
\caption{(a) In SA module, a set of neighbor points are grouped and followed by MLPs and max pooling for feature aggregation. (b) In residual SA module, SA is firstly used to extract the local feature for the centroid point, and then the residual connection is applied by adding input and output features. The linear layer is not necessary when $C$ is equal to $C'$.}
\label{fig:res_sa}
\end{figure}

\vspace{3pt}
\noindent
\textbf{Atom pointset.} We denote $\mathcal{A}$ as the atom pointset with $M_1$ points, where $i^\text{th}$ atom point $(\textbf{p}_{a_i}, \textbf{f}_{a_i}) \in \mathbb{R}^3 \times \mathbb{R}^6$ consists of a 3-dim XYZ coordinate vector $\textbf{p}_{a_i}$ to indicate the 3D spatial location, and a 6-dim one-hot vector $\textbf{f}_{a_i}$ to represent the atom type (i.e., C, H, O, N, S, Se).

\begin{figure*}[t]
\centering
\includegraphics[width=1.0\linewidth]{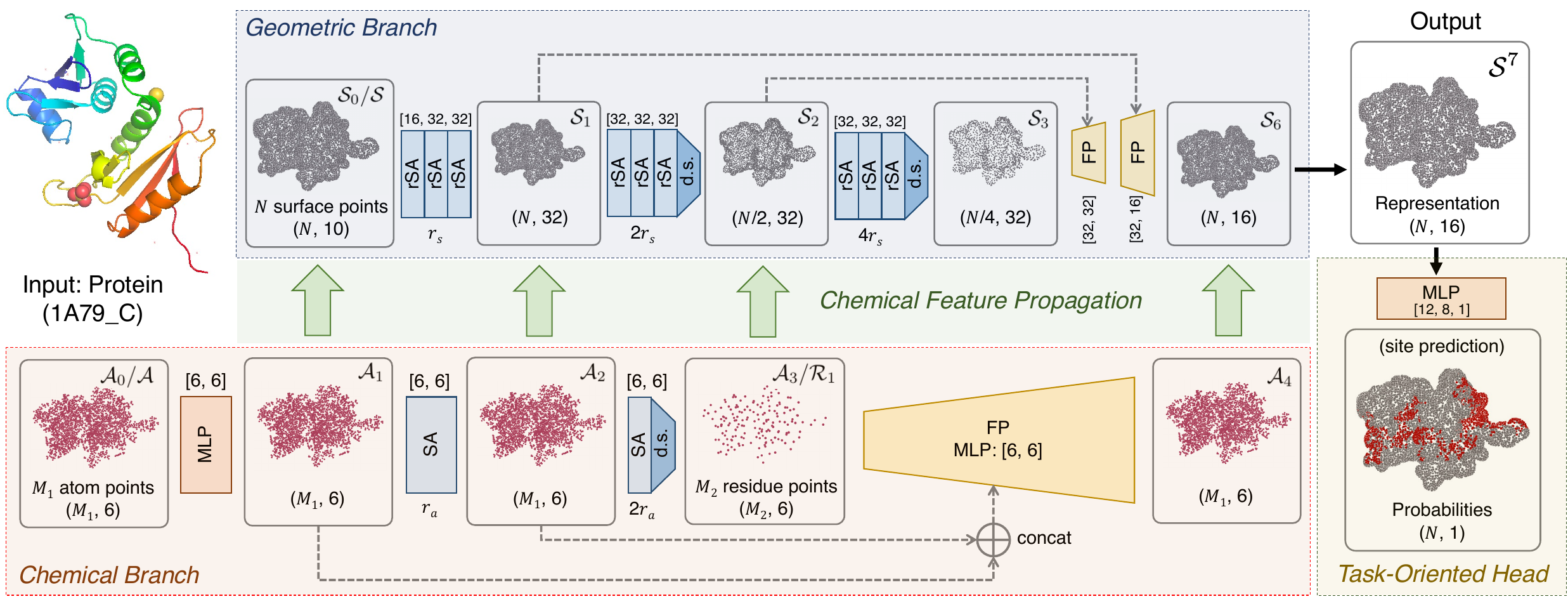}
\vspace{-6mm}
\caption{\ud{The backbone of \nickname{}} is mainly composed of two branches for geometric (top) and chemical (bottom) feature learning. Two branches encode features from surface points and atom points in a multiscale way, respectively. Chemical feature propagation modules (middle) are proposed to propagate features from the chemical branch to the geometric branch also in a hierarchical way. Moreover, different task-oriented heads can be followed to handle different downstream tasks, such as site prediction and interaction matching. ``d.s.'' indicates point cloud downsampling. $r_s$ and $r_a$ are the initial query radii for two branches, respectively.}
\label{fig:architecture}
\end{figure*}

\vspace{3pt}
\noindent
\textbf{Residue pointset.} Suppose that the protein is composed of $M_2$ amino acid residues, where $i^\text{th}$ residue is formed by $N_i$ atoms. Regarding the indices of atoms are denoted as $\{d_1, \dots, d_{N_i}\}$, the center position of the residue is calculated by averaging 3-dim coordinates of all internal atoms:
\begin{equation}
    \textbf{p}_{r_i} = \frac{1}{N_i} \sum_{k=1}^{N_i} \textbf{p}_{a_{d_k}}.
\end{equation}
Hence, the residue pointset is represented by $\mathcal{R}=\big\{\textbf{p}_{r_i}\big\}_{i=1}^{M_2} \subset \mathbb{R}^3$, which is used as the downsampled pointset to aggregate features from atom points for hierarchical chemical feature learning, which will be described in Section~\ref{sec:backbone}.

\vspace{3pt}
\noindent
\textbf{Surface pointset.} To perform effective learning on the protein surface, we follow \cite{dmasif_cvpr2021} to generate $N (N \approx 2.5M_1)$ surface points from atoms with a fast sampling algorithm. Let $\mathcal{S}$ be the surface pointset, where $i^\text{th}$ surface point $(\textbf{p}_{s_i}, \textbf{f}_{s_i}) \in \mathbb{R}^3 \times \mathbb{R}^{10}$ contains a 3-dim XYZ coordinate vector $\textbf{p}_{s_i}$ and a 10-dim feature vector $\textbf{f}_{s_i}$. $\textbf{f}_{s_i}$ are Gaussian and mean curvatures estimated at 5 scales ranging from 1\r{A} to 10\r{A}.

\subsection{Preliminaries of Point Cloud Processing} \label{sec:preliminary}
\noindent
As mentioned, a protein can be represented by three pointsets, which can be processed in the way of point cloud learning. PointNet++~\cite{qi2017pointnetplusplus} is one of the most popular methods in point cloud processing and has been widely used in many computer vision tasks such as shape classification and scene understanding. It consists of two main modules, set abstraction (SA) module to encode features and downscale points and feature propagation (FP) module to upscale points and decode features. These two modules and the modified version are used as basic point (de)convolution operators in our framework.

\vspace{3pt}
\noindent
\textbf{Set abstraction (SA) module} groups neighbor points around a centroid point and then uses a mini-PointNet~\cite{qi2016pointnet} to encode local patterns into feature vectors.
%
%
Let $\mathcal{P} = \big\{(\textbf{p}_i, \textbf{f}_i)\big\}_{i=1}^{N_p} \subset \mathbb{R}^3 \times \mathbb{R}^C$ be a pointset with $N_p$ points and $C$ refers to the dimension of the feature vector. Given a centroid point $(\textbf{p}', \textbf{f}')$,
%
%
use ball query to group neighbor points' features as 
$
    \mathcal{N}(\textbf{p}', \mathcal{P}) = \Big\{\big[\textbf{p} - \textbf{p}'; \textbf{f}\big]\ \big\lvert\ \lVert\textbf{p} - \textbf{p}'\rVert_2 \leq r, (\textbf{p}, \textbf{f}) \in \mathcal{P} \big\}
$ 
with $r \in \mathbb{R}$ being the chosen radius and $[\cdot;\cdot]$ is the concatenation operator.
Then we apply shared multi-layer perceptrons (MLPs) followed by a max pooling layer on $\mathcal{N}(\textbf{p}', \mathcal{P})$ and output the local feature $\hat{\textbf{f}'}$ at $\textbf{p}'$:
\begin{equation}
\begin{split}
    \hat{\textbf{f}'} &= \text{SA}(\textbf{p}', \mathcal{P}) \\
                        &= \text{MAX}\Big\{\text{MLP}(\hat{\textbf{f}})\  \Big\lvert\  \hat{\textbf{f}} = \big[\textbf{p} - \textbf{p}'; \textbf{f}\big] \in \mathcal{N}(\textbf{p}', \mathcal{P}) \Big\}.
\end{split}
\end{equation}
As shown in Figure~\ref{fig:res_sa}, to build a deep learning network, we develop a residual version of SA (rSA) inspired by \cite{he2016deep} by adding the original feature $\textbf{f}'$ at $\textbf{p}'$:
\begin{equation}
\hat{\textbf{f}'} = \text{rSA}(\textbf{p}', \textbf{f}', \mathcal{P}) 
                    = \text{SA}(\textbf{p}', \mathcal{P}) + \textbf{f}'.
\end{equation}

\vspace{3pt}
\noindent
\textbf{Feature propagation (FP) module.} Points are downsampled in an encoder-decoder network, and FP module is used to recover the data size and propagate features from downsampled pointset $\mathcal{P}'$ to the original pointset $\mathcal{P}$. To be more specific, for each point $(\textbf{p}, \textbf{f}) \in \mathcal{P}$, we first select $K$ nearest neighbor points from $\mathcal{P}'$ as $\mathcal{N}_K(\textbf{p}, \mathcal{P}') = \big\{(\textbf{p}_{k}', \textbf{f}_{k}')\big\}_{k=1}^K$. Then we interpolate feature values using their relative distances:
\begin{equation}
    \textbf{f}_\text{interp} = \text{Interp}(\textbf{p}, \mathcal{N}_K(\textbf{p}, \mathcal{P}')) =  \frac{\sum_k w_k \textbf{f}_{k}'}{\sum_k w_k},
\end{equation}
where $w_k = 1/\lVert\textbf{p}_{k}' - \textbf{p}\rVert_2 \in \mathbb{R}$ is the interpolation weight. Furthermore, we adopt skip connection to combine propagated features $\textbf{f}_\text{interp}$ and original features $\textbf{f}$ together and use a shared MLP to update each point's feature:
\begin{equation}
    \hat{\textbf{f}} = \text{FP}(\textbf{p}, \textbf{f}, \mathcal{P}')
                                = \text{MLP}\left([\textbf{f}_\text{interp}; \textbf{f}]\right).
\end{equation}

\subsection{Hierarchical Feature Interaction Network} \label{sec:backbone}
\noindent
As addressed in Section~\ref{sec:introduction}, the multiscale relationship among atoms and hierarchical feature interactions between surface points and atoms are significant and overlooked by previous works. In this section, we formally propose \nickname{} to model the above hierarchical relationships. An overview of the network architecture is shown in Figure~\ref{fig:architecture}. Given a protein represented by three pointsets (atom, residue, and surface), we feed surface points into the geometric branch to encode hierarchical geometry-related features and atom/residue points into the chemical branch to encode hierarchical chemistry-related features. More importantly, the chemical feature propagation module is introduced in a hierarchical way to enhance the feature interaction between two branches to improve feature representation learning. Finally, we obtain the representative features of each surface point and design task-oriented heads for different protein surface learning tasks. 

\vspace{3pt}
\noindent
\textbf{Geometric branch} is composed of three encoding layers and two decoding (upsampling) layers to learn geometric features from surface points in a hierarchical way, as shown at the top of Figure~\ref{fig:architecture}. To formulate, in $i^\text{th}$ encoding layer, assume the input is $\mathcal{S}^i \in \mathbb{R}^{N_i \times C_i}$ with $N_i$ points and a channel size of $C_i$. A stack of rSA modules (see Section~\ref{sec:preliminary}) are applied for local feature encoding. The channels of MLP in $k^\text{th}$ rSA module are $[C_i^{k-1}, C_i^k, C_i^k, C_i^k]$, where $C_i^{k-1}$ and $C_i^k$ are the output channel sizes of $k-1^\text{th}$ and $k^\text{th}$ rSA modules respectively, and $C_i^0 = C_i$. All rSA modules in $i^\text{th}$ encoding layer have the same neighbor query radius $r_i$, which is twice the radius $r_{i-1}$ of the previous encoding layer (i.e., $r_i = 2r_{i-1}$). Let  $r_1 = r_s$ be the initial radius. In some encoding layers, farthest point sampling (FPS~\cite{eldar1997farthest}) is adopted following the final rSA module to downsample half of surface points for efficient learning. In decoding layers, FP modules are used to propagate features from downsampled pointset to the original pointset layer by layer. Skip connections are used to combine low-level and high-level features to obtain a better representation. The initial radius $r_s$ is set to 4\r{A} in site prediction and 2\r{A} in interaction matching. The output channel size of each module is given in Figure~\ref{fig:architecture}.

\begin{figure}[t]
\centering
\includegraphics[width=1.0\linewidth]{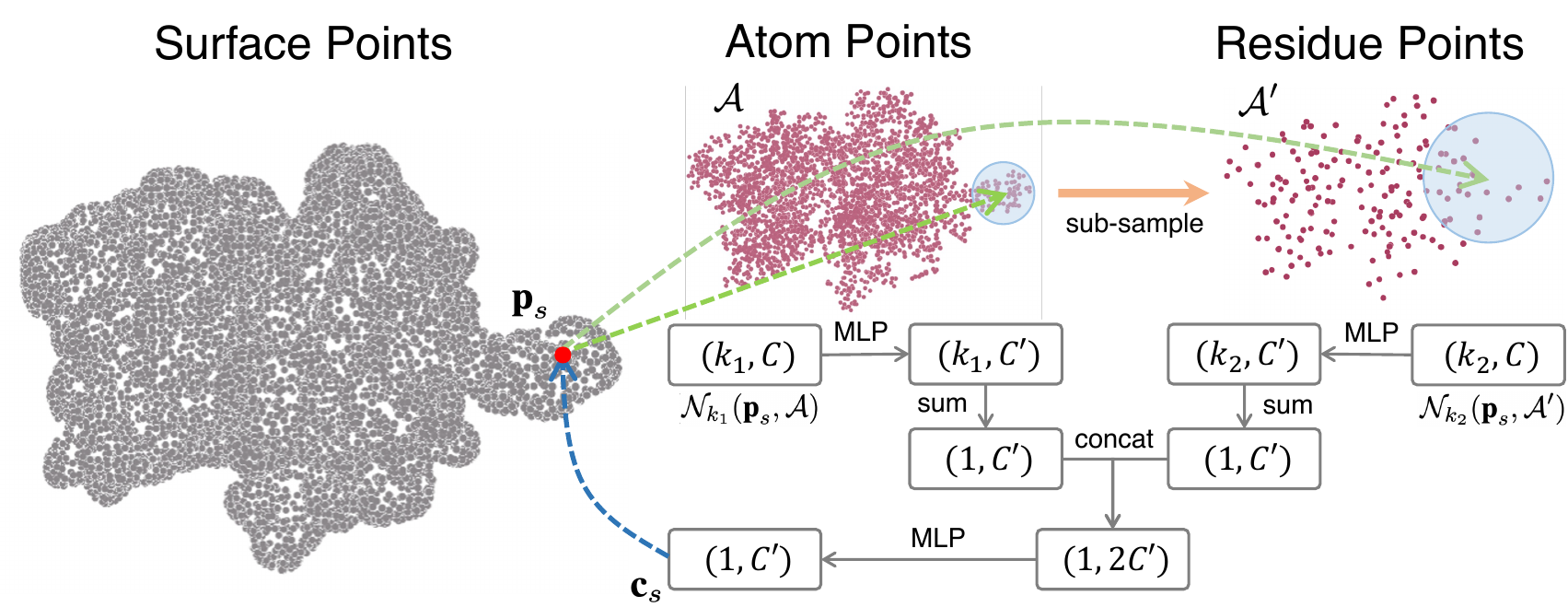}
\vspace{-5mm}
\caption{Implementation details of feature aggregation function $g(\cdot)$ in CFP. $g(\cdot)$ operates on each surface point to query chemical features from neighbor atoms (near) and residues (far). We use MLPs to transform features and summation to aggregate features from neighbors.}
\label{fig:chem_interp}
\end{figure}

\vspace{3pt}
\noindent
\textbf{Chemical branch} is composed of three encoding layers and one decoding layer for multiscale chemical feature learning, as shown at the bottom of Figure~\ref{fig:architecture}. Different from the geometric branch, the first encoding layer ($\mathcal{A}_1$) is a shared MLP to encode atom features individually from one-hot atom type vectors. In the rest of encoding layers ($\mathcal{A}_2$ and $\mathcal{R}_1$), the neighbor query radii are $r_a$ and $2r_a$, where $r_a$ is the initial radius and set to 4\r{A} in site prediction and 3\r{A} in interaction matching. We use SA modules without the residual connection for local feature encoding since the network is not as deep as the geometric branch. 
Instead of using FPS to downsample the pointset, residue points $\mathcal{R}=\big\{\textbf{p}_{r_i}\big\}_{i}$ are used as centroids to group neighbor atom points and obtain $\mathcal{R}_1 = \{(\textbf{p}_{r_i}, \textbf{f}_{r_i}^1)\}_i$:
\begin{equation}
    \textbf{f}_{r_i}^1 = \text{SA}(\textbf{p}_{r_i}, \mathcal{A}_2),
\end{equation}
where $\mathcal{A}_2$ are output atom points of the first SA module and $\textbf{f}_{r_i}^1$ is the aggregated feature vector at residue $\textbf{p}_{r_i}$. Then the FP module and skip connection are used to combine all levels of chemical features in the upsampling layer ($\mathcal{A}_4$). Hence, we have chemical features in four levels: individual, local, global, and mixed, from $\mathcal{A}_1$ to $\mathcal{A}_4$ (let $\mathcal{A}_3 = \mathcal{R}_1$).

\vspace{3pt}
\noindent
\textbf{Chemical feature propagation (CFP) module} is introduced to propagate features from chemical branch to geometric branch. More specifically, to propagate features from $\mathcal{A}$ to each surface point $(\textbf{p}_{s}, \textbf{f}_{s}) \in \mathcal{S}$ (subscripts of $\mathcal{A}$ and $\mathcal{S}$ are ignored for simplification), we denote
\begin{equation}
\begin{split}
    \hat{\textbf{f}}_{s} &= \text{CFP}(\textbf{p}_{s}, \textbf{f}_{s}, \mathcal{A})
                            = h(\textbf{c}_{s}, \textbf{f}_{s}),\ \text{and} \\
    \textbf{c}_{s}       &= g\left(\mathcal{N}_{k_1}(\textbf{p}_{s}, \mathcal{A}), \mathcal{N}_{k_2}(\textbf{p}_{s}, \mathcal{A}')\right),
\end{split}
\end{equation}
where $\mathcal{N}_{k_1/k_2}(\cdot)$ are kNN selection functions with $k$ equal to $k_1$ and $k_2$, respectively. 
$\mathcal{A}'$ is the sparse pointset sub-sampled from $\mathcal{A}$ and used for querying far away chemical points.
$g(\cdot)$ is a feature aggregation function operated on chemical pointsets, $\textbf{c}_{s}$ is the propagated chemical feature, $h(\cdot)$ is a multi-modality fusion function to combine chemical and geometric features, and $\hat{\textbf{f}}_{s}$ is the fused feature. 
In practice, we regard residue points as the sub-sampled centroids and assign each residue feature by the feature of the nearest point in $\mathcal{A}$ to form $\mathcal{A}'$. If the chemical source is the residue set (e.g., $\mathcal{A}_3\rightarrow \mathcal{S}_2$), then $\mathcal{A}' = \mathcal{A}$. The detailed design of $g(\cdot)$ is shown in Figure~\ref{fig:chem_interp}, and we directly concatenate the propagated chemical features and the original geometric features together:
\begin{equation}
    \hat{\textbf{f}}_{s} = h(\textbf{c}_{s}, \textbf{f}_{s}) = [\textbf{c}_{s}; \textbf{f}_{s}].
\end{equation}
As shown in the middle of Figure~\ref{fig:architecture}, we adopt CFP modules to bridge the chemical points (atoms or residues) and surface points that have similar receptive fields to establish hierarchical interactions between chemical and geometric branches.

\subsection{Task-Oriented Heads for Downstream Tasks}
\noindent
The outputs of the backbone are representative point-wise features on the protein surface, which can be formulated as
\begin{equation}
    \mathcal{S}^7 = \big\{(\textbf{p}_{s_i}, \textbf{f}_{s_i}^7)\big\}_{i=1}^N 
                  = \text{CFP}(\mathcal{S}_6, \mathcal{A}_4),
\end{equation}
where $\mathcal{S}_6$ and $\mathcal{A}_4$ are the output pointsets of geometric and chemical branches, respectively. Hence we can design task-oriented heads to handle different tasks. In this work, two tasks are included, interface site prediction and interaction matching~\cite{masif_nature_method,dmasif_cvpr2021}.

\vspace{3pt}
\noindent
\textbf{Site prediction} aims to identify interaction sites and non-interaction sites from surface points. As shown in Figure~\ref{fig:architecture}, a three-layer MLP is followed to predict the probability of being an interaction site for each surface point $(\textbf{p}_{s}, \textbf{f}_{s}^7) \in \mathcal{S}^7$. The full model is end-to-end optimized with binary cross entropy (BCE) loss and Dice loss~\cite{milletari2016v}:
\begin{equation}
\begin{split}
    \mathcal{L}(\hat{y}_{s}, y_{s}) &= \lambda \mathcal{L}_\text{BCE}(\hat{y}_{s}, y_{s}) + \\
                                        &\ \ \ \ (1 - \lambda) \mathcal{L}_\text{Dice}(\hat{y}_{s}, y_{s}),\ \text{and} \\
    \hat{y}_{s} &= \text{MLP}(\textbf{f}_{s}^7),
\end{split}
\label{eq:site}
\end{equation}
where $\lambda \in [0, 1]$ is trade-off coefficient, $\hat{y}_{s} \in \mathbb{R}$ is final predicted probability, and $y_{s} \in \{0, 1\}$ is ground truth label.

\begin{figure*}[t]
\centering
\includegraphics[width=0.95\linewidth]{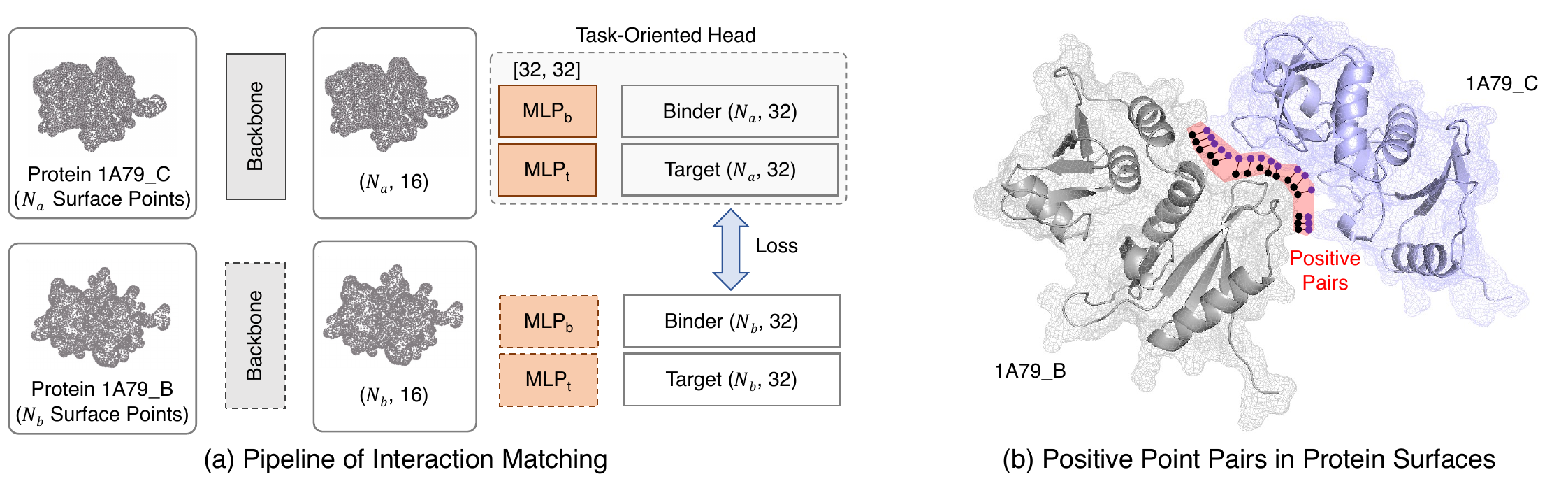}
\vspace{-3mm}
\caption{(a) Illustration of interaction matching network. The backbone network is shown in Figure~\ref{fig:architecture}. The task-oriented head comprises two MLPs to predict the representative feature (binder) and corresponding complementary feature (target). (b) An example: point pairs (colored in red) located very close in the formed complex are labeled as positive pairs.}
\label{fig:matching}
\end{figure*}

\vspace{3pt}
\noindent
\textbf{Interaction matching} aims to predict the interaction probability of each surface point pair, one from each protein involved in a complex, which is key to protein docking. Because we are about to measure feature complementarity instead of similarity, as shown in Figure~\ref{fig:matching}.a, two branches are used to predict the representative feature (binder) and the corresponding complemental feature (target), respectively. To formulate, for each surface point $(\textbf{p}_{s}, \textbf{f}_{s}^7) \in \mathcal{S}^7$, the output binder feature $\textbf{b}$ and target feature $\textbf{t}$ are
\begin{equation}
    \textbf{b} = \text{MLP}_b(\textbf{f}_{s}^7)
    \ \text{and} \ 
    \textbf{t} = \text{MLP}_t(\textbf{f}_{s}^7),
\end{equation}
where $\text{MLP}_b$ and $\text{MLP}_t$ are MLP layers of binder and target branches. 
%
%
During training, the input is a pair of proteins (denoted as A and B, both are randomly rotated) from the same complex. The backbone network first produces binder and target features for the surface points of each protein.
Given point $a$ from protein A with output features $(\textbf{t}_a, \textbf{b}_a)$ and point $b$ from protein B with $(\textbf{t}_b, \textbf{b}_b)$, the interaction probability of these two points is formulated as
\begin{equation}
    p(a, b) = \text{Sigmoid}\left(\frac{\textbf{t}_{a}^T \textbf{b}_{b} + \textbf{t}_{b}^T \textbf{b}_{a}}{2}\right).
\end{equation}
If the relative distance between point $a$ and point $b$ is less than 1\r{A} before the random rotation of proteins, the point pair $(a, b)$ is labeled as a positive pair; otherwise, it is labeled as a negative pair. Therefore, the objective function is defined as
\begin{equation}
    \mathcal{L}(a, b, y_{ab}) = \mathcal{L}_{\text{BCE}}\big(p(a, b), y_{ab}\big),
    \label{eq:matching}
\end{equation}
where $y_{ab} \in \{0, 1\}$ is 1 if and only if $(a, b)$ is a positive pair. All positive pairs are selected, and the same number of negative pairs are randomly sampled for training. Due to the memory limitation, gradients are only recorded during the forward pass of the first protein (i.e., \textit{1A79\_C} in Figure~\ref{fig:matching}.a).

\section{Experiments}
\noindent
To validate the effectiveness of the proposed HCGNet, we conduct experiments on two tasks in the field of structural bioinformatics, which are introduced in \cite{masif_nature_method} and aim to deal with interactions between proteins. We implement our method with PyTorch~\cite{paszke2017automatic}. Momentum gradient descent optimizer is used to optimize loss function in Eqn.~\ref{eq:site} and Eqn.~\ref{eq:matching} with an initial learning rate of 0.01 and a batch size of 8. Leaky ReLU and instance normalization are applied in MLP layers. Models are trained on a single NVIDIA GeForce RTX 3090 GPU for 300 epochs ($\sim$24h) in site prediction and 150 epochs ($\sim$36h) in interaction matching.

\subsection{Site Prediction}

\noindent\textbf{Setting.}
In this task, the surface points of a protein are classified into interaction sites and non-interaction sites. Interaction sites are surface points that are more likely to interact with other proteins. This task can benefit protein engineering, such as drug discovery. We use the dataset proposed in \cite{masif_nature_method}, which contains 3,314 proteins collected from the Protein Data Bank~\cite{berman2003announcing} with 2,958 proteins for training and 356 for testing. We follow \cite{masif_nature_method} to split 10\% data from the training set for validation during training \ud{based on the pairwise matrix of TM-scores (refer to Supplementary Note 4 in \cite{masif_nature_method} for more details)}. $\lambda$ in Eqn.~\ref{eq:site} is set to 0.5. To quantitatively compare our proposed \nickname{} with previous works, we follow \cite{dmasif_cvpr2021} to report ROC-AUC (area under the ROC curve) as the evaluation metric, \ud{since it measures the ability to distinguish between classes and is insensitive to class imbalance.}

\begin{figure*}[t]
\centering
\includegraphics[width=1.0\linewidth]{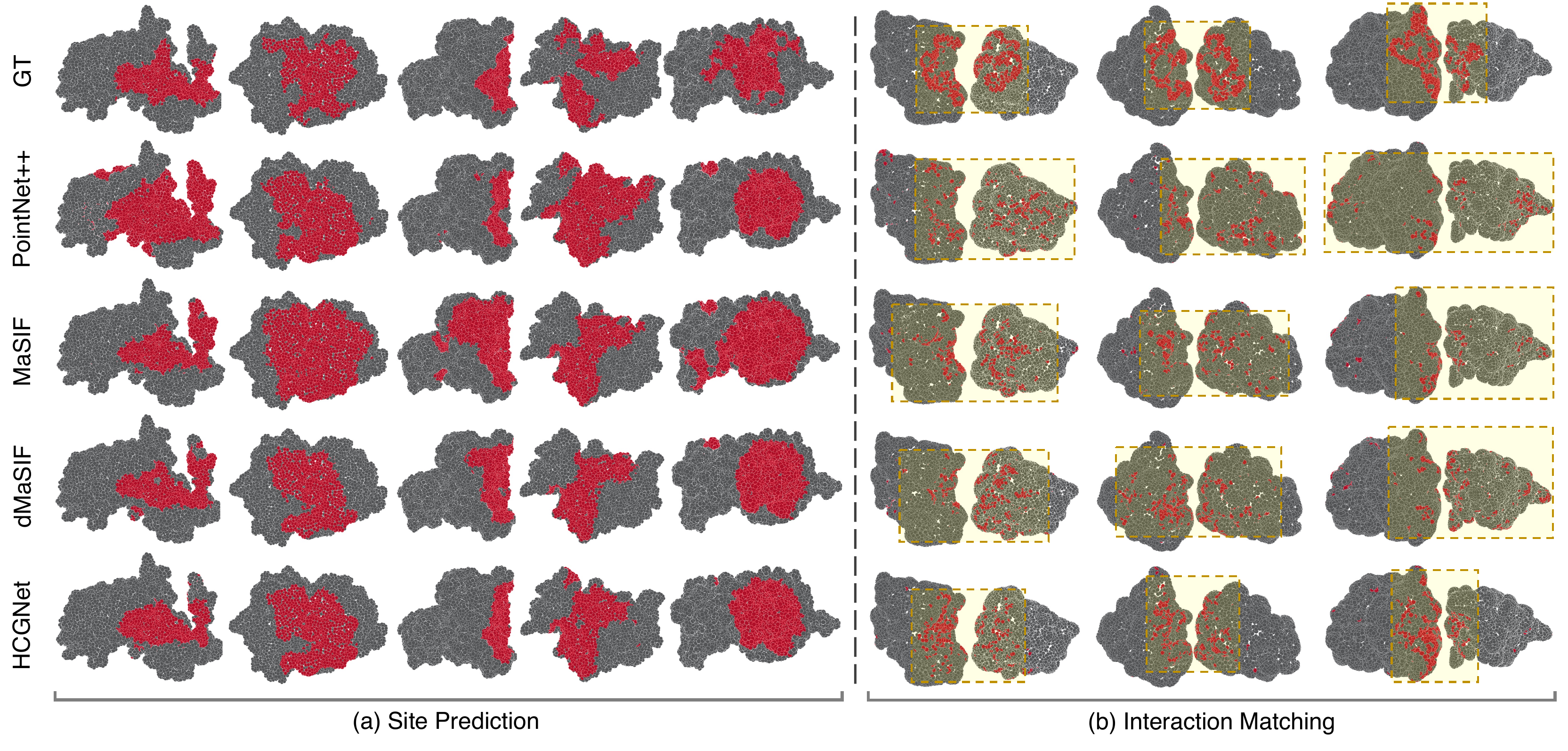}
\vspace{-7.5mm}
\caption{\ud{Qualitative results on site prediction and interaction matching tasks.} (a) Our HCGNet can perform more precise segmentation of interaction sites (red) than \ud{previous methods}. (b) Matched points are highlighted in red. Compared with \ud{previous methods}, our predicted point pairs can be more helpful for the orientation alignment since most matched points are located in associating areas; see bounded areas in yellow.}
\label{fig:vis_all}
\end{figure*}

\vspace{3pt}
\noindent\textbf{Results.}
As shown in Table~\ref{tab:all_res}, \nickname{} outperforms the SoTA method (dMaSIF~\cite{dmasif_cvpr2021}; see also Figure~\ref{fig:roc}.a) by 2.3\% ROC-AUC and is significantly ahead of other previous methods. In addition, we visualize the results in Figure~\ref{fig:vis_all}.a. Compared with \ud{previous methods}, our predicted interaction sites are more precise with the help of hierarchical chemical features.

\begin{table}[t]
\centering
\setlength{\tabcolsep}{14pt}
\caption{Area under the ROC curve (ROC-AUC) is evaluated to compare our proposed architecture with previous methods on site prediction and interaction matching.}
\vspace{-1mm}
\resizebox{1.0\columnwidth}{!}{
\begin{tabular}{l|c|cc}
\toprule[1.2pt]
Method & \begin{tabular}[c]{@{}c@{}}Site \\ Prediction\end{tabular} & \begin{tabular}[c]{@{}c@{}}Interaction \\ Matching\end{tabular} \\ \hline \hline
DGCNN~\cite{dgcnn} & 0.710 & - \\
\ud{PointNet~\cite{qi2016pointnet}} & \ud{0.816} & \ud{0.747} \\
PointNet++~\cite{qi2017pointnetplusplus} & 0.848 & 0.785 \\
\ud{PointConv~\cite{wu2018pointconv}} & \ud{0.855} & \ud{0.790} \\
MaSIF~\cite{masif_nature_method} & 0.850 & 0.787 \\
dMaSIF~\cite{dmasif_cvpr2021} & 0.870 & 0.794 \\
 \hline
HCGNet (\textit{ours}) & \textbf{0.893} & \textbf{0.826} \\
\bottomrule[1.2pt]
\end{tabular}
}
\label{tab:all_res}
\end{table}

\subsection{Interaction Matching}
\noindent\textbf{Setting.}
In this task, two proteins involved in a complex are given to predict the interaction probability of point pairs (one from each). This task is the key to protein docking, which aims to align the orientation of two proteins in a complex. We use the dataset proposed in \cite{masif_nature_method}, including 4,614 protein pairs for training and 912 for testing. 10\% training data are used for validation \ud{based structural alignments (refer to Supplementary Note 6 in \cite{masif_nature_method} for more details)}. In the evaluation stage, all positive pairs are selected, and the same number of negative pairs are randomly sampled for ROC-AUC calculation. 

\begin{figure}[t]
\centering
\includegraphics[width=1.0\linewidth]{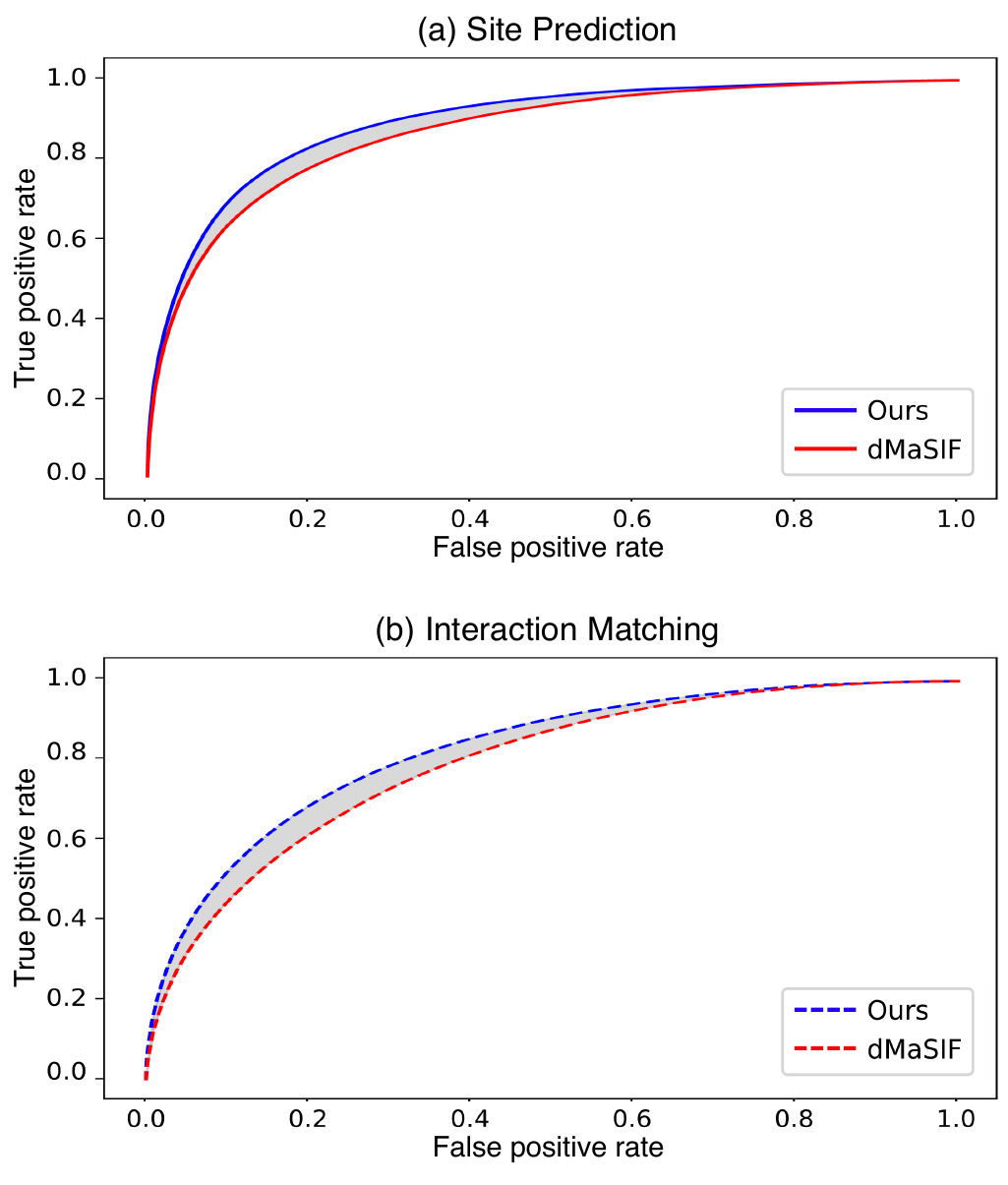}
\vspace{-6mm}
\caption{ROC curves comparing the performance of our method (blue) and dMaSIF (red)~\cite{dmasif_cvpr2021} on the task of site prediction (a) and interaction matching (b).}
\label{fig:roc}
\end{figure}

\vspace{3pt}
\noindent\textbf{Results.}
As quantitative comparison shown in Table~\ref{tab:all_res}, our \nickname{} performs better than the SoTA (dMaSIF~\cite{dmasif_cvpr2021}; see also Figure~\ref{fig:roc}.b) and all other previous methods with a remarkable margin. Note that dMaSIF~\cite{dmasif_cvpr2021} reported a ROC-AUC of 0.82 on interaction matching by selecting all positive pairs and 400 negative pairs for evaluation. Based on this evaluation strategy, our performance (0.837) is still better. In Figure~\ref{fig:vis_all}.b, we highlight the matched points in red. Visual results show that our predicted pairs are more concentrated in the associating areas. 
Compared to site prediction, our proposed architecture has achieved more improvements in interaction matching (see also the second ablation study). The reason may be that interaction matching is aware of binding partners, and hierarchical chemical features can help the network to identify specific chemical complementarities such as alkaline and acidic, positively charged, and negatively charged. However, site prediction aims to predict general interaction sites where the functional group is active and the geometric shape is obviously concave or convex.

\subsection{Ablation Study}
\noindent
To further analyze our proposed architecture, ablative experiments are conducted to explore the importance of chemical and geometric features, the importance of different proposed parts, and different designs of hierarchical learning networks and hierarchical feature interaction. 
%

\vspace{6pt}
\noindent
\textbf{Importance of chemical and geometric features.}
To illustrate the significance of chemical and geometric features, we design two backbones: \textit{chem.~only} to use only the chemical branch and propagate features in the final layer to surface points for point-wise classification, and \textit{geom.~only} to use only the geometric branch without any chemical feature propagation. Experiments are conducted on the tasks of site prediction and interaction matching. As shown in Table~\ref{tab:geom_chem}, in the absence of either chemical features or geometric features, the performance drops dramatically in both two tasks, which validates the importance of the interaction between chemical features and geometric features.

\begin{table}[t]
\centering
\setlength{\tabcolsep}{15pt}
\caption{The importance of chemical and geometric features. ROC-AUC is evaluated and the performance drops sharply when training with only chemical (chem.) features or only geometric (geom.) features in both two tasks.}
\vspace{-1mm}
\resizebox{1.0\columnwidth}{!}{
\begin{tabular}{l|c|c}
\toprule[1.2pt]
Method & \begin{tabular}[c]{@{}c@{}}Site \\ Prediction\end{tabular} & \begin{tabular}[c]{@{}c@{}}Interaction\\ Matching\end{tabular} \\ \hline \hline
chem. only & 0.673 & 0.692 \\
geom. only & 0.759 & 0.722 \\ \hline
chem. \& geom. & \textbf{0.893} & \textbf{0.826} \\ 
\bottomrule[1.2pt]
\end{tabular}
}
\label{tab:geom_chem}
\end{table}

\vspace{6pt}
\noindent
\textbf{Importance of different proposed parts.}
To demonstrate the effectiveness of different proposed parts, we conduct ablative experiments by removing CFP, hierarchical chemical feature learning, and hierarchical interaction between two branches, respectively. In Table~\ref{tab:cmp_backbones}, the performance drops without any of the above parts, which implies the significance of modeling multiscale relationships among atoms and hierarchical feature interaction between surface points and atoms (or residues). In addition, we notice that by removing hierarchical chemical learning and CFP, the performance drops for interaction matching (-1.3\%/-2.3\%) are more than site prediction (-0.9\%/-1.9\%) since awareness of binding partner requires more precise chemical features to predict chemical complementarity.

\begin{table}[t]
\centering
\caption{The importance of chemical feature learning and hierarchical feature interaction. ROC-AUC is evaluated for comparing the full model with removing different proposed parts, including CFP, hierarchical chemical (chem.) feature learning, and hierarchical feature interaction (inter.) between two branches.}
\vspace{-1mm}
\setlength{\tabcolsep}{10pt}
\resizebox{1.0\columnwidth}{!}{
\begin{tabular}{l|c|c}
\toprule[1.2pt]
Method & \begin{tabular}[c]{@{}c@{}}Site \\ Prediction\end{tabular} & \begin{tabular}[c]{@{}c@{}}Interaction\\ Matching\end{tabular} \\ \hline \hline
w/o CFP & \multicolumn{1}{l|}{0.874 \textcolor{red}{\scriptsize -1.9\%} } & \multicolumn{1}{l}{0.803 \textcolor{red}{\scriptsize -2.3\%}} \\
w/o hierarchical chem. & \multicolumn{1}{l|}{0.884 \textcolor{red}{\scriptsize -0.9\%}} & \multicolumn{1}{l}{0.813 \textcolor{red}{\scriptsize -1.3\%}} \\
w/o hierarchical inter. & \multicolumn{1}{l|}{0.886 \textcolor{red}{\scriptsize -0.7\%} } & \multicolumn{1}{l}{0.820 \textcolor{red}{\scriptsize -0.6\%}} \\ \hline
Full model (\nickname{}) & \multicolumn{1}{l|}{\textbf{0.893}} & \multicolumn{1}{l}{\textbf{0.826}} \\
\bottomrule[1.2pt]
\end{tabular}
}
\label{tab:cmp_backbones}
\end{table}

\vspace{6pt}
\noindent
\textbf{Design on hierarchical branches.}
As mentioned, hierarchical analysis is significant for both chemical and geometric learning. We compare different radii of ball query in SA (or rSA) modules from 2\r{A} to 7\r{A} in two tasks, as shown in Table~\ref{tab:com_radius}. We find that the best setting has similar initial radii in two learning branches, $(r_s, r_a) = (4, 4)$\r{A} in site prediction and $(r_s, r_a) = (2, 3)$\r{A} in interaction matching.

\begin{table}[t]
\centering
\caption{Comparison of different initial radii $r_s$ in the geometric branch and $r_a$ the chemical branch. ROC-AUC is evaluated, and the best setting has similar $r_s$ and $r_a$ in both two tasks.}
\vspace{-1mm}
\setlength{\tabcolsep}{13pt}
\resizebox{1.0\columnwidth}{!}{
\begin{tabular}{cc|c|c}
\toprule[1.2pt]
$r_s$ (\r{A}) & $r_a$ (\r{A}) & \begin{tabular}[c]{@{}c@{}}Site \\ Prediction\end{tabular} & \begin{tabular}[c]{@{}c@{}}Interaction\\ Matching\end{tabular} \\ \hline \hline
2 & 3 & 0.889 & \textbf{0.826} \\
4 & 2 & 0.888 & 0.814 \\
4 & 4 & \textbf{0.893} & 0.825  \\
4 & 7 & 0.887 & 0.821 \\
7 & 4 & 0.883 & 0.813 \\
\bottomrule[1.2pt]
\end{tabular}
}
\label{tab:com_radius}
\end{table}

\vspace{6pt}
\noindent
\textbf{Design on hierarchical feature interaction.}
\ud{In \nickname{}, the feature propagation from the chemical branch to the geometric branch is implemented as $(\mathcal{A}_1, \mathcal{A}_2, \mathcal{A}_3, \mathcal{A}_4)$$\rightarrow$$(\mathcal{S}_0, \mathcal{S}_1, \mathcal{S}_2, \mathcal{S}_6)$, which indicates $\mathcal{A}_1$$\rightarrow$$\mathcal{S}_0$, $\mathcal{A}_2$$\rightarrow$$\mathcal{S}_1$, $\mathcal{A}_3$$\rightarrow$$\mathcal{S}_2$, and $\mathcal{A}_4$$\rightarrow$$\mathcal{S}_6$.}
\ud{In} \ud{Table~\ref{tab:cfp_link}, we compare different chemical sources for the propagation process.}
Although the feature propagation between two branches can be arbitrary, we still recommend establishing the interaction in the way of Figure~\ref{fig:architecture}, because the linked surface points and chemical points (atoms or residues) have similar receptive fields. If not, the network may be confused by inconsistent receptive fields during training and bring worse performance. 

\begin{table}[t]
\centering
\caption{Comparison of different propagation strategies. ROC-AUC is evaluated, and the chemical sources represent the atom pointset propagated to $\mathcal{S}_0$, $\mathcal{S}_1$, $\mathcal{S}_2$ and $\mathcal{S}_6$, respectively. - indicates no chemical feature propagation to this surface pointset.}
\vspace{-1mm}
\setlength{\tabcolsep}{13pt}
\resizebox{1.0\columnwidth}{!}{
\begin{tabular}{l|c|c}
\toprule[1.2pt]
Chemical Sources & \begin{tabular}[c]{@{}c@{}}Site \\ Prediction\end{tabular} & \begin{tabular}[c]{@{}c@{}}Interaction\\ Matching\end{tabular} \\ \hline \hline
($\mathcal{A}_1$, -, -, -) & 0.884 & 0.813 \\
($\mathcal{A}_4$, $\mathcal{A}_3$, $\mathcal{A}_2$, $\mathcal{A}_1$) & 0.888 & 0.818 \\
($\mathcal{A}_1$, $\mathcal{A}_3$, $\mathcal{A}_3$, $\mathcal{A}_4$) & 0.887 & 0.820 \\
($\mathcal{A}_1$, $\mathcal{A}_2$, $\mathcal{A}_3$, $\mathcal{A}_4$) & \textbf{0.893} & \textbf{0.826} \\
\bottomrule[1.2pt]
\end{tabular}
}
\label{tab:cfp_link}
\end{table}

\vfill


\section{Discussion}

\subsection{Biomolecular Surface Learning}
\noindent
Biomolecules, including proteins, are composed of atoms and most chemical interactions occur on biomolecular surfaces. This means the surface features (e.g., chemistry, geometry) play an important role in interaction-related tasks. In this work, we point out that the relationship among atoms and the feature propagation from atoms to surfaces are significant for protein surface learning, which should be the same for other biomolecules' surface learning. More importantly, we propose a general framework \nickname{} mainly focusing on the design of chemical learning and feature propagation. \nickname{} can handle different downstream tasks, such as site prediction and interaction matching, by equipping different task-oriented heads. We believe that \nickname{} can also handle tasks in other biomolecules, such as DNA/RNA and ligands, with only minor modifications to the proposed framework. The remaining difficulties are the task definition and dataset construction. We look forward to and will work towards the unification of frameworks for biomolecular surface learning.

\begin{table}[t]
\centering
\setlength{\tabcolsep}{8pt}
\caption{Comparison of model parameters. We compare the model parameters of training only with chemical features (chem. only), training only with geometric features (geom. only), removing CFP (w/o CFP), removing hierarchical chemical learning (w/o hierarchical chem.), and removing hierarchical feature interaction (w/o hierarchical inter.).}
\vspace{-1mm}
\resizebox{1.0\columnwidth}{!}{
\begin{tabular}{l||c|c||c|c}
\toprule[1.2pt]
Method & \multicolumn{2}{c||}{\begin{tabular}[c]{@{}c@{}}Site\\ Prediction\end{tabular}} & \multicolumn{2}{c}{\begin{tabular}[c]{@{}c@{}}Interaction\\ Matching\end{tabular}} \\ \hline
chem. only & 0.673 & 0.67K & 0.692 & 5.08K \\
geom. only & 0.759 & 24.4K & 0.722 & 29.3K \\ \hline
w/o CFP    & 0.874 & 27.6K & 0.803 & 32.9K \\
w/o hierarchical chem.  & 0.884 & 27.3K & 0.813 & 32.5K \\
w/o hierarchical inter. & 0.886 & 24.9K & 0.820 & 29.9K \\ \hline
Full model (\nickname{}) & \textbf{0.893} & 28.3K & \textbf{0.826} & 33.6K \\ 
\bottomrule[1.2pt]
\end{tabular}
}
\label{tab:param}
\end{table}

\subsection{Processing Efficiency}
\noindent
Compared with dMaSIF~\cite{dmasif_cvpr2021}, \nickname{} has more parameters (28.3K vs. 2.8K) and requires more GPU memories (407MB/protein vs. 132MB/protein) due to the design of deep network. 
However, due to the hierarchical design, the inference time of \nickname{} (17ms) is comparable with dMaSIF~\cite{dmasif_cvpr2021} (16ms), \ud{and much less than other compared methods\footnote{For the inference time of DGCNN, PointNet++, MaSIF, and dMaSIF, we report the results presented in \cite{dmasif_cvpr2021}.}, including DGCNN~\cite{dgcnn} (50ms), PointNet~\cite{qi2016pointnet} (130ms), PointNet++~\cite{qi2017pointnetplusplus} (410ms), PointConv~\cite{wu2018pointconv} (45ms), and MaSIF~\cite{masif_nature_method} (180ms).}
In addition, we should point out that 28.3K is not a big number in deep learning networks (e.g., $\sim$12M in ResNet18), 
and our proposed CFP, hierarchical chemical learning and interaction indeed do not introduce many training parameters but bring considerable improvements; see Section~\ref{sec:params}.
All the above results are tested under the same environment setting of Ubuntu 20.04, RTX 3090, CUDA 11.1, and PyTorch 1.8.


\subsection{Model Parameters} \label{sec:params}
\noindent
In Table~\ref{tab:param}, we compare the performance and training parameters of networks in the ablative experiments, 
which shows that introducing CFP, hierarchical chemical learning and interaction can bring significant improvements (1.9\%/0.9\%/0.7\%; 2.3\%/1.3\%/0.6\%) 
with few increments of training parameters (0.7K/1.0K/3.4K; 0.7K/1.1K/3.7K) in both two tasks.

\subsection{Scalability}

\noindent
\ud{ The proposed framework is scalable to some extent. When processing larger molecules or more complicated downstream tasks, it would be better to increase the model capability by increasing the depth/width of chemical and geometric branches. However, increased computational requirements and overfitting could be possible issues when doing so.}

\section{Conclusion}
\noindent
In this work, we highlight the importance of the multiscale relationship between atoms and the hierarchical interaction between chemical and geometric features. To this end, we propose \nickname{}, a novel learning architecture for protein surface analysis. \nickname{} takes atoms and surface points of a given protein as the input. Then two hierarchical branches are used to learn chemical features from atoms and geometric features from surface points in parallel. In addition, features are hierarchically propagated from the chemical branch to the geometric branch for multi-modality feature fusion. Our experiments demonstrate that \nickname{} significantly improves the performance over the SoTA method in two challenging protein analysis tasks: site prediction and interaction matching.

\vspace{3pt}
\noindent
\ud{\textbf{Limitations and future work.} Although \nickname{} is a general framework for joint learning from chemical and geometric features, the hyperparameters and the model design should be specifically tuned for different downstream applications. Hence, it is more important to develop a unified framework and collect numerous data for general biomolecular surface learning, which will left as our future work.}

\vfill

\section{Acknowledgement}
\noindent This work is partially supported by grants from the National Natural Science Foundation of China (No. 62306254) and the Foshan HKUST Projects under Grants FSUST21-HKUST10E and FSUST21-HKUST11E. This research is also supported under the RIE2020 Industry Alignment Fund - Industry Collaboration Projects (IAF-ICP) Funding Initiative, as well as cash and in-kind contributions from the industry partner(s).

\bibliographystyle{IEEEtran}
\bibliography{egbib}

\begin{thebibliography}{10}
\providecommand{\url}[1]{#1}
\csname url@samestyle\endcsname
\providecommand{\newblock}{\relax}
\providecommand{\bibinfo}[2]{#2}
\providecommand{\BIBentrySTDinterwordspacing}{\spaceskip=0pt\relax}
\providecommand{\BIBentryALTinterwordstretchfactor}{4}
\providecommand{\BIBentryALTinterwordspacing}{\spaceskip=\fontdimen2\font plus
\BIBentryALTinterwordstretchfactor\fontdimen3\font minus
  \fontdimen4\font\relax}
\providecommand{\BIBforeignlanguage}[2]{{%
\expandafter\ifx\csname l@#1\endcsname\relax
\typeout{** WARNING: IEEEtran.bst: No hyphenation pattern has been}%
\typeout{** loaded for the language `#1'. Using the pattern for}%
\typeout{** the default language instead.}%
\else
\language=\csname l@#1\endcsname
\fi
#2}}
\providecommand{\BIBdecl}{\relax}
\BIBdecl

\bibitem{shen2020machine}
C.~Shen, J.~Ding, Z.~Wang, D.~Cao, X.~Ding, and T.~Hou, ``From machine learning
  to deep learning: Advances in scoring functions for protein--ligand
  docking,'' \emph{Wiley Interdisciplinary Reviews: Computational Molecular
  Science}, vol.~10, no.~1, p. e1429, 2020.

\bibitem{wang2002investigation}
Y.~Wang and O.~Jardetzky, ``Investigation of the neighboring residue effects on
  protein chemical shifts,'' \emph{Journal of the American Chemical Society},
  vol. 124, no.~47, pp. 14\,075--14\,084, 2002.

\bibitem{jones1996principles}
S.~Jones and J.~M. Thornton, ``Principles of protein-protein interactions,''
  \emph{Proceedings of the National Academy of Sciences}, vol.~93, no.~1, pp.
  13--20, 1996.

\bibitem{novotny1992electrostatic}
J.~Novotny and K.~Sharp, ``Electrostatic fields in antibodies and
  antibody/antigen complexes,'' \emph{Progress in biophysics and molecular
  biology}, vol.~58, no.~3, pp. 203--224, 1992.

\bibitem{roberts1991electrostatic}
V.~A. Roberts, H.~Freeman, A.~Olson, J.~Tainer, and E.~Getzoff, ``Electrostatic
  orientation of the electron-transfer complex between plastocyanin and
  cytochrome c,'' \emph{Journal of Biological Chemistry}, vol. 266, no.~20, pp.
  13\,431--13\,441, 1991.

\bibitem{braden1995structural}
B.~C. Braden and R.~J. Poljak, ``Structural features of the reactions between
  antibodies and protein antigens,'' \emph{The FASEB Journal}, vol.~9, no.~1,
  pp. 9--16, 1995.

\bibitem{demchuk1994receptor}
E.~Demchuk, T.~Mueller, H.~Oschkinat, W.~Sebald, and R.~C. Wade, ``Receptor
  binding properties of four-helix-bundle growth factors deduced from
  electrostatic analysis,'' \emph{Protein Science}, vol.~3, no.~6, pp.
  920--935, 1994.

\bibitem{vakser1994hydrophobic}
I.~A. Vakser and C.~Aflalo, ``Hydrophobic docking: a proposed enhancement to
  molecular recognition techniques,'' \emph{Proteins: Structure, Function, and
  Bioinformatics}, vol.~20, no.~4, pp. 320--329, 1994.

\bibitem{lee1971interpretation}
B.~Lee and F.~M. Richards, ``The interpretation of protein structures:
  estimation of static accessibility,'' \emph{Journal of molecular biology},
  vol.~55, no.~3, pp. 379--IN4, 1971.

\bibitem{lawrence1993shape}
M.~C. Lawrence and P.~M. Colman, ``Shape complementarity at protein/protein
  interfaces,'' 1993.

\bibitem{masif_nature_method}
P.~Gainza, F.~Sverrisson, F.~Monti, E.~Rodol{\`a}, D.~Boscaini, M.~Bronstein,
  and B.~Correia, ``Deciphering interaction fingerprints from protein molecular
  surfaces using geometric deep learning,'' \emph{Nature Methods}, vol.~17,
  no.~2, pp. 184--192, 2020.

\bibitem{mai2020multiscale}
G.~Mai, K.~Janowicz, B.~Yan, R.~Zhu, L.~Cai, and N.~Lao, ``Multi-scale
  representation learning for spatial feature distributions using grid cells,''
  2020.

\bibitem{rego2021identifying}
N.~B. Rego, E.~Xi, and A.~J. Patel, ``Identifying hydrophobic protein patches
  to inform protein interaction interfaces,'' \emph{Proceedings of the National
  Academy of Sciences}, vol. 118, no.~6, 2021.

\bibitem{dmasif_cvpr2021}
F.~Sverrisson, J.~Feydy, B.~E. Correia, and M.~M. Bronstein, ``Fast end-to-end
  learning on protein surfaces,'' in \emph{The IEEE Conference on Computer
  Vision and Pattern Recognition (CVPR)}, June 2021.

\bibitem{mcnaught1997compendium}
A.~D. McNaught, A.~Wilkinson \emph{et~al.}, \emph{Compendium of chemical
  terminology}.\hskip 1em plus 0.5em minus 0.4em\relax Blackwell Science
  Oxford, 1997, vol. 1669.

\bibitem{AlphaFold2021}
J.~Jumper, R.~Evans, A.~Pritzel, T.~Green, M.~Figurnov, O.~Ronneberger,
  K.~Tunyasuvunakool, R.~Bates, A.~{\v{Z}}{\'\i}dek, A.~Potapenko,
  A.~Bridgland, C.~Meyer, S.~A.~A. Kohl, A.~J. Ballard, A.~Cowie,
  B.~Romera-Paredes, S.~Nikolov, R.~Jain, J.~Adler, T.~Back, S.~Petersen,
  D.~Reiman, E.~Clancy, M.~Zielinski, M.~Steinegger, M.~Pacholska,
  T.~Berghammer, S.~Bodenstein, D.~Silver, O.~Vinyals, A.~W. Senior,
  K.~Kavukcuoglu, P.~Kohli, and D.~Hassabis, ``Highly accurate protein
  structure prediction with {AlphaFold},'' \emph{Nature}, vol. 596, no. 7873,
  pp. 583--589, 2021.

\bibitem{senior2019protein}
A.~W. Senior, R.~Evans, J.~Jumper, J.~Kirkpatrick, L.~Sifre, T.~Green, C.~Qin,
  A.~{\v{Z}}{\'\i}dek, A.~W. Nelson, A.~Bridgland \emph{et~al.}, ``Protein
  structure prediction using multiple deep neural networks in the 13th critical
  assessment of protein structure prediction (casp13),'' \emph{Proteins:
  Structure, Function, and Bioinformatics}, vol.~87, no.~12, pp. 1141--1148,
  2019.

\bibitem{senior2020improved}
A.~W. Senior, R.~Evans, Jumper \emph{et~al.}, ``Improved protein structure
  prediction using potentials from deep learning,'' \emph{Nature}, vol. 577,
  no. 7792, pp. 706--710, 2020.

\bibitem{xu2019distance}
J.~Xu, ``Distance-based protein folding powered by deep learning,''
  \emph{Proceedings of the National Academy of Sciences}, vol. 116, no.~34, pp.
  16\,856--16\,865, 2019.

\bibitem{dai2021protein}
B.~Dai and C.~Bailey-Kellogg, ``Protein interaction interface region prediction
  by geometric deep learning,'' \emph{Bioinformatics}, 2021.

\bibitem{jamasb2021deep}
A.~R. Jamasb, B.~Day, C.~Cangea, P.~Li{\`o}, and T.~L. Blundell, ``Deep
  learning for protein--protein interaction site prediction,'' in
  \emph{Proteomics Data Analysis}.\hskip 1em plus 0.5em minus 0.4em\relax
  Springer, 2021, pp. 263--288.

\bibitem{somnath2021multi}
V.~R. Somnath, C.~Bunne, and A.~Krause, ``Multi-scale representation learning
  on proteins,'' \emph{Advances in Neural Information Processing Systems},
  vol.~34, 2021.

\bibitem{xiong2021featurization}
G.~Xiong, C.~Shen, Z.~Yang, D.~Jiang, S.~Liu, A.~Lu, X.~Chen, T.~Hou, and
  D.~Cao, ``Featurization strategies for protein--ligand interactions and their
  applications in scoring function development,'' \emph{Wiley Interdisciplinary
  Reviews: Computational Molecular Science}, p. e1567, 2021.

\bibitem{pierce2011accelerating}
B.~G. Pierce, Y.~Hourai, and Z.~Weng, ``Accelerating protein docking in zdock
  using an advanced 3d convolution library,'' \emph{PloS one}, vol.~6, no.~9,
  p. e24657, 2011.

\bibitem{renaud2021deeprank}
N.~Renaud, C.~Geng, S.~Georgievska, F.~Ambrosetti, L.~Ridder, D.~F. Marzella,
  A.~M. Bonvin, and L.~C. Xue, ``Deeprank: A deep learning framework for data
  mining 3d protein-protein interfaces,'' \emph{Biorxiv}, 2021.

\bibitem{schneidman2005patchdock}
D.~Schneidman-Duhovny, Y.~Inbar, R.~Nussinov, and H.~J. Wolfson, ``Patchdock
  and symmdock: servers for rigid and symmetric docking,'' \emph{Nucleic acids
  research}, vol.~33, no. suppl\_2, pp. W363--W367, 2005.

\bibitem{mikolov2013efficient}
T.~Mikolov, K.~Chen, G.~Corrado, and J.~Dean, ``Efficient estimation of word
  representations in vector space,'' \emph{arXiv preprint arXiv:1301.3781},
  2013.

\bibitem{le2014distributed}
Q.~Le and T.~Mikolov, ``Distributed representations of sentences and
  documents,'' in \emph{International conference on machine learning}.\hskip
  1em plus 0.5em minus 0.4em\relax PMLR, 2014, pp. 1188--1196.

\bibitem{alley2019unified}
E.~C. Alley, G.~Khimulya, S.~Biswas, M.~AlQuraishi, and G.~M. Church, ``Unified
  rational protein engineering with sequence-based deep representation
  learning,'' \emph{Nature methods}, vol.~16, no.~12, pp. 1315--1322, 2019.

\bibitem{krause2017multiplicative}
B.~Krause, L.~Lu, I.~Murray, and S.~Renals, ``Multiplicative lstm for sequence
  modelling,'' 2017.

\bibitem{simonovsky2020deeplytough}
M.~Simonovsky and J.~Meyers, ``Deeplytough: learning structural comparison of
  protein binding sites,'' \emph{Journal of chemical information and modeling},
  vol.~60, no.~4, pp. 2356--2366, 2020.

\bibitem{townshend2019endtoend}
R.~J.~L. Townshend, R.~Bedi, P.~A. Suriana, and R.~O. Dror, ``End-to-end
  learning on 3d protein structure for interface prediction,'' 2019.

\bibitem{fout2017protein}
F.~A., B.~J., S.~B., and B.-H. A., ``Protein interface prediction using graph
  convolutional networks,'' in \emph{Advances in Neural Information Processing
  System}, 2017.

\bibitem{qi2016pointnet}
C.~R. Qi, H.~Su, K.~Mo, and L.~J. Guibas, ``Pointnet: Deep learning on point
  sets for 3d classification and segmentation,'' \emph{arXiv preprint
  arXiv:1612.00593}, 2016.

\bibitem{qi2017pointnetplusplus}
C.~R. Qi, L.~Yi, H.~Su, and L.~J. Guibas, ``Pointnet++: Deep hierarchical
  feature learning on point sets in a metric space,'' \emph{arXiv preprint
  arXiv:1706.02413}, 2017.

\bibitem{li2018pointcnn}
Y.~Li, R.~Bu, M.~Sun, W.~Wu, X.~Di, and B.~Chen, ``Pointcnn: Convolution on
  x-transformed points,'' \emph{Advances in neural information processing
  systems}, vol.~31, pp. 820--830, 2018.

\bibitem{thomas2019KPConv}
H.~Thomas, C.~R. Qi, J.-E. Deschaud, B.~Marcotegui, F.~Goulette, and L.~J.
  Guibas, ``Kpconv: Flexible and deformable convolution for point clouds,''
  \emph{Proceedings of the IEEE International Conference on Computer Vision},
  2019.

\bibitem{wu2018pointconv}
W.~Wu, Z.~Qi, and L.~Fuxin, ``Pointconv: Deep convolutional networks on 3d
  point clouds,'' \emph{arXiv preprint arXiv:1811.07246}, 2018.

\bibitem{li2019deepgcns}
G.~Li, M.~Muller, A.~Thabet, and B.~Ghanem, ``Deepgcns: Can gcns go as deep as
  cnns?'' in \emph{Proceedings of the IEEE/CVF International Conference on
  Computer Vision}, 2019, pp. 9267--9276.

\bibitem{dgcnn}
Y.~Wang, Y.~Sun, Z.~Liu, S.~E. Sarma, M.~M. Bronstein, and J.~M. Solomon,
  ``Dynamic graph cnn for learning on point clouds,'' \emph{ACM Transactions on
  Graphics (TOG)}, 2019.

\bibitem{tatarchenko2018tangent}
M.~Tatarchenko, J.~Park, V.~Koltun, and Q.-Y. Zhou, ``Tangent convolutions for
  dense prediction in 3d,'' in \emph{Proceedings of the IEEE Conference on
  Computer Vision and Pattern Recognition}, 2018, pp. 3887--3896.

\bibitem{lin2020fpconv}
Y.~Lin, Z.~Yan, H.~Huang, D.~Du, L.~Liu, S.~Cui, and X.~Han, ``Fpconv: Learning
  local flattening for point convolution,'' in \emph{Proceedings of the
  IEEE/CVF Conference on Computer Vision and Pattern Recognition}, 2020, pp.
  4293--4302.

\bibitem{berman2003announcing}
H.~Berman, K.~Henrick, and H.~Nakamura, ``Announcing the worldwide protein data
  bank,'' \emph{Nature Structural \& Molecular Biology}, vol.~10, no.~12, pp.
  980--980, 2003.

\bibitem{he2016deep}
K.~He, X.~Zhang, S.~Ren, and J.~Sun, ``Deep residual learning for image
  recognition,'' in \emph{Proceedings of the IEEE conference on computer vision
  and pattern recognition}, 2016, pp. 770--778.

\bibitem{eldar1997farthest}
Y.~Eldar, M.~Lindenbaum, M.~Porat, and Y.~Y. Zeevi, ``The farthest point
  strategy for progressive image sampling,'' \emph{IEEE Transactions on Image
  Processing}, vol.~6, no.~9, pp. 1305--1315, 1997.

\bibitem{milletari2016v}
F.~Milletari, N.~Navab, and S.-A. Ahmadi, ``V-net: Fully convolutional neural
  networks for volumetric medical image segmentation,'' in \emph{2016 fourth
  international conference on 3D vision (3DV)}.\hskip 1em plus 0.5em minus
  0.4em\relax IEEE, 2016, pp. 565--571.

\bibitem{paszke2017automatic}
A.~Paszke, S.~Gross, S.~Chintala, G.~Chanan, E.~Yang, Z.~DeVito, Z.~Lin,
  A.~Desmaison, L.~Antiga, and A.~Lerer, ``Automatic differentiation in
  pytorch,'' in \emph{NIPS-W}, 2017.

\end{thebibliography}

\end{document}